\documentclass{ws-procs9x6}

\usepackage{graphics}
\usepackage{subfigure}

\def\pT{\mbox{$p_T$}}

\def\sqrtsNN{\mbox{$\sqrt{s_\mathrm{_{NN}}}$}}
\def\sqrts{\mbox{$\sqrt{s}$}}
\def\Npart{\mbox{$\mathrm{N}_\mathrm{part}$}}
\def\NpartMean{\mbox{$\langle\Npart\rangle$}}
\def\Nbinary{\mbox{$\mathrm{N}_\mathrm{bin}$}}
\def\NbinaryMean{\mbox{$\langle\Nbinary\rangle$}}

\def\hplus{\mbox{$\mathrm{h}^+$}}
\def\hminus{\mbox{$\mathrm{h}^-$}}
\def\hphm{\mbox{$(\hplus+\hminus)/2$}}

\def\RAA{\mbox{$R_{AA}(\pT)$}}

\def\TAA{\mbox{$T_{AA}$}}

\def\sigmaNNinel{\mbox{$\sigma^{NN}_{inel}$}}
\def\sigmaAAgeom{\mbox{$\sigma^{AuAu}_{geom}$}}

\def\lt{\mbox{$<$}}

\begin{document}

\title{Overview of results from the STAR experiment at RHIC\footnote{\uppercase{T}his work was supported by the \uppercase{D}ivision of \uppercase{N}uclear 
\uppercase{P}hysics and the \uppercase{D}ivision of \uppercase{H}igh \uppercase{E}nergy \uppercase{P}hysics of the \uppercase{O}ffice of \uppercase{S}cience of 
the \uppercase{U}.\uppercase{S}. \uppercase{D}epartment of \uppercase{E}nergy, the \uppercase{U}nited \uppercase{S}tates \uppercase{N}ational \uppercase{S}cience \uppercase{F}oundation,
the \uppercase{B}undesministerium f\"ur \uppercase{B}ildung und \uppercase{F}orschung of \uppercase{G}ermany,
the \uppercase{I}nstitut \uppercase{N}ational de la \uppercase{P}hysique \uppercase{N}ucl\'eaire et de la \uppercase{P}hysique 
des \uppercase{P}articules of \uppercase{F}rance, the \uppercase{U}nited \uppercase{K}ingdom \uppercase{E}ngineering and \uppercase{P}hysical 
\uppercase{S}ciences \uppercase{R}esearch \uppercase{C}ouncil, \uppercase{F}unda\c c\~ao de \uppercase{A}mparo \`a \uppercase{P}esquisa do \uppercase{E}stado de \uppercase{S}\~ao \uppercase{P}aulo, \uppercase{B}razil, the \uppercase{R}ussian \uppercase{M}inistry of \uppercase{S}cience and \uppercase{T}echnology and the
\uppercase{M}inistry of \uppercase{E}ducation of \uppercase{C}hina and the \uppercase{N}ational \uppercase{S}cience \uppercase{F}oundation of \uppercase{C}hina.
}}\author{K. FILIMONOV for the STAR collaboration}

\address{Nuclear Science Division, \\
Lawrence Berkeley National Laboratory, \\ 
1 Cyclotron Road,
Berkeley, CA 94720, USA\\ 
E-mail: KVFilimonov@lbl.gov}


\maketitle

\abstracts{The Relativistic Heavy-Ion Collider (RHIC) provides
Au+Au collisions at energies up to \sqrtsNN=200 GeV. 
STAR experiment was designed and constructed to investigate 
the behavior of strongly interacting matter at high energy density.
An overview of some of the recent
results from the STAR collaboration is given.}

\section{Introduction}
One of the fundamental predictions of the 
Quantum Chromodynamics (QCD) is the existence of a deconfined 
state of quarks and gluons at the energy densities above 
1 GeV/fm$^3$~\cite{Harris:1996zx}.
This strongly interacting medium, the Quark Gluon Plasma (QGP), 
may be created in the laboratory by the collision of heavy nuclei
at high energy. The current experimental program
at the Relativistic Heavy-Ion Collider (RHIC) is aimed at detecting
the new state of matter and studying its properties. 

The Relativistic Heavy-Ion Collider is located at Brookhaven
National Laboratory (New York, USA). It is capable of colliding
gold ions from  \sqrtsNN=20 to 200 GeV per nucleon pair, protons up to
\sqrts=500 GeV, and asymmetric systems like deuterons with 
heavy nuclei. RHIC is also the first polarized proton collider at high energies, opening
new opportunities to study the spin structure of the proton.
Diffractive processes in high electromagnetic fields can also
be studied in ultra-peripheral heavy-ion collisions.

\section{STAR Detector}
\begin{figure}
\centering
\mbox{
\subfigure{\includegraphics[height=0.53\textwidth]{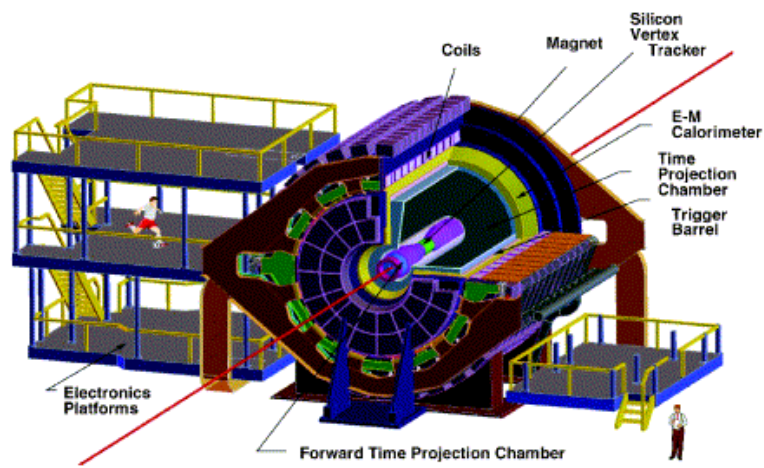}}
}
\caption{Schematic view of the STAR detector.
\label{star}}
\end{figure}

The Solenoidal Tracker at RHIC (STAR) is one of two large 
detector systems constructed at RHIC \cite{nim}. 
The layout of the STAR experiment
is shown in Figure~\ref{star}. 
The main detector in STAR is the world's largest Time Projection Chamber
(TPC) \cite{Anderson:2003ur} measuring trajectories of charged particles at mid-rapidity ($|\eta|<1.4$) with full azimuthal coverage. 
A solenoidal magnet
provides a homogeneous magnetic field up to 0.5~T. 
Charged particle tracking close to the interaction region is 
accomplished by a Silicon Vertex Tracker consisting of
three layers of silicon drift detectors. The high
pseudorapidity range ($2.5<|\eta|<4$) is covered by two 
radial drift Forward TPCs.
A barrel electromagnetic calorimeter (EMC) has been partially installed.
Eventually, the full-barrel EMC and an end-cap electromagnetic calorimeter
will have a combined coverage of $-1<|\eta|<2$.
Particles are identified by their specific energy loss in the TPC gas.
Particle decays in the TPC volume are identified by either their decay
topology or on a statistical basis by reconstructing the invariant mass
of daughter candidates. 

\section{Soft Physics}

Relativistic heavy-ion collisions produce a large number of 
hadrons, most of which are in the ``soft'' low transverse momentum
\pT$<2$ GeV/c region of phase space. 
These hadrons carry important information about the collision dynamics
from the early stage of the collisions to the final state interactions.
The particle ratios are sensitive to the chemical
properties of the system and the particle production mechanism.
Measurements of the particle multiplicity, pseudorapidity, and transverse
momentum distributions are valuable tools to study the bulk properties
of the created system and its evolution. 

\subsection{Baryon Stopping and Chemical Freeze-out}
Baryon number transport (or stopping), achieved at early stages
of high-energy collisions, is an important observable sensitive to
the overall dynamical evolution of the collisions.
The primordial QGP to hadron phase transition a few $10^{-6}$ 
seconds after the Big Bang occurred in a nearly net-baryon free region.
A dramatic increase in the antibaryon-to-baryon ratios measured
at mid-rapidity is observed
from SPS ($\bar{p}/p\approx 0.07$) to RHIC ($\bar{p}/p\approx 0.7$)\cite{Adler:2001bp,Adler:2002pba} energies, 
indicating that the system created at RHIC is getting close to
net-baryon free.

\begin{figure}[htbp]
\centerline{\epsfxsize=4.4in\epsfysize=2in\epsfbox{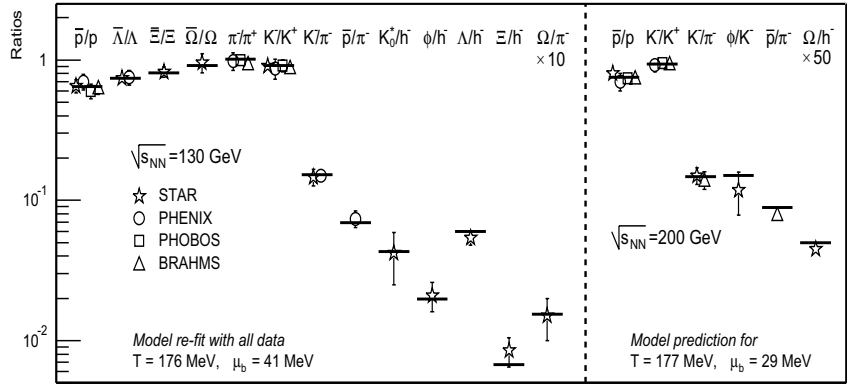}}
\caption{Particle  ratios measured  by the four  RHIC collaborations
are  compared to the statistical model calculations.
}
\label{fig5}
\end{figure}

STAR is particularly well-suited for studying the particle production,
with results obtained for $\pi^0$, $\pi^-$, $\pi^+$, $K^-$, $K^+$, $K^0_s$, 
$\rho$, $K^{*0}$+$\overline{K^{*0}}$, f$_0$, $p$, $\bar{p}$, $\phi$,
$\Lambda$, $\overline{\Lambda}$, $\Xi^-$, $\overline{\Xi}^+$, $\Omega^-$, and
$\overline{\Omega}^+$~\cite{VanBuren:2002sp}. 
One can attempt to describe the chemical
composition of the system within a thermal, chemical equilibrium model.
From the experimentally measured particle ratios
the chemical freeze-out parameters can be extracted.  Chemical freeze-out 
occurs when flavor changing inelastic 
collisions cease.  At this point the yields of various
particles are fixed, and subsequent elastic scattering will not change the 
particle composition.  Since the cross-sections for inelastic collisions are
smaller than the cross-sections for elastic collisions, one expects the 
chemical ratios to be fixed at a time before kinetic freeze-out.  
Figure~\ref{fig5} (taken from~\cite{Ullrich:2002tq})  
shows the particle ratios measured at mid-rapidity
for a wide variety of non-strange and strange hadrons, with the results
from a statistical model fit to the data~\cite{Braun-Munzinger:2001ip}.
 The resulting fit from this and similar 
models is excellent,
yielding a chemical freeze-out temperature and baryon chemical potential of
$T_{ch}\approx 175$ MeV and $\mu_{b}\approx 25-50$ MeV, close to
the critical temperature for the deconfinement phase transition on the lattice.

\subsection{Particle Spectra and Kinetic Freeze-out}

\begin{figure}[htb]
\centerline{
\epsfxsize=0.28\textwidth\epsfbox[110 240 430 580]{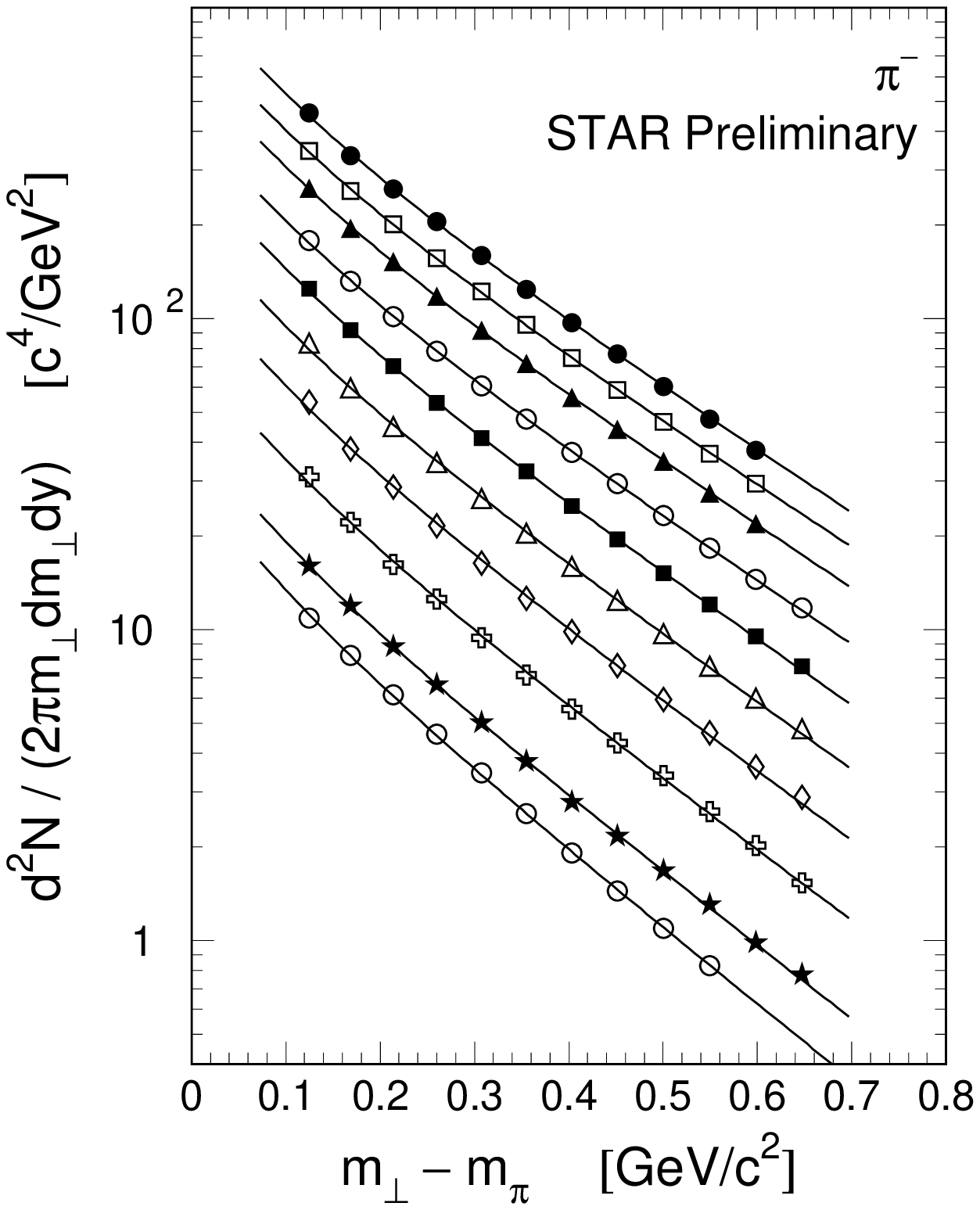}
\epsfxsize=0.28\textwidth\epsfbox[110 240 430 580]{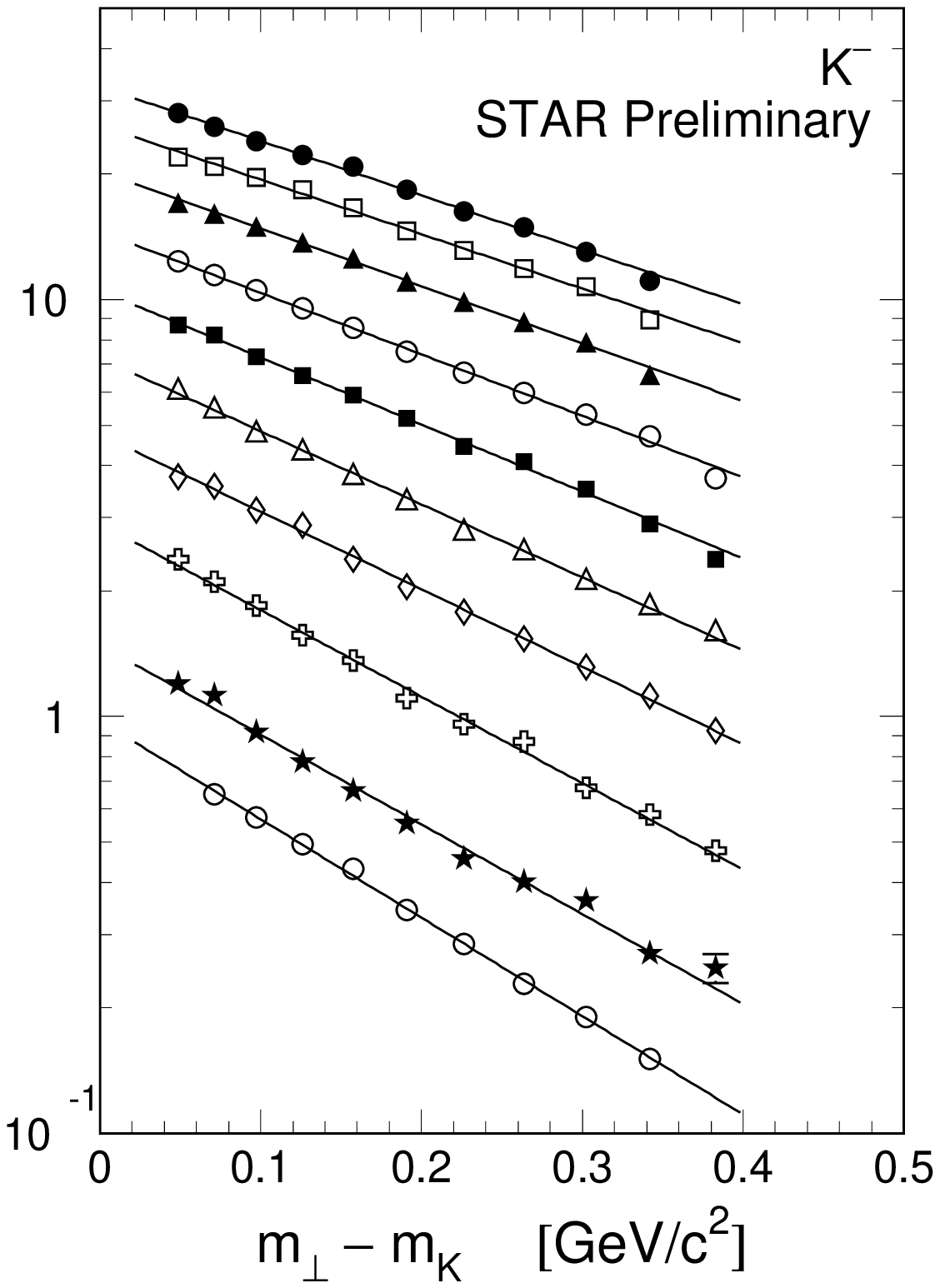}
\epsfxsize=0.28\textwidth\epsfbox[110 240 430 580]{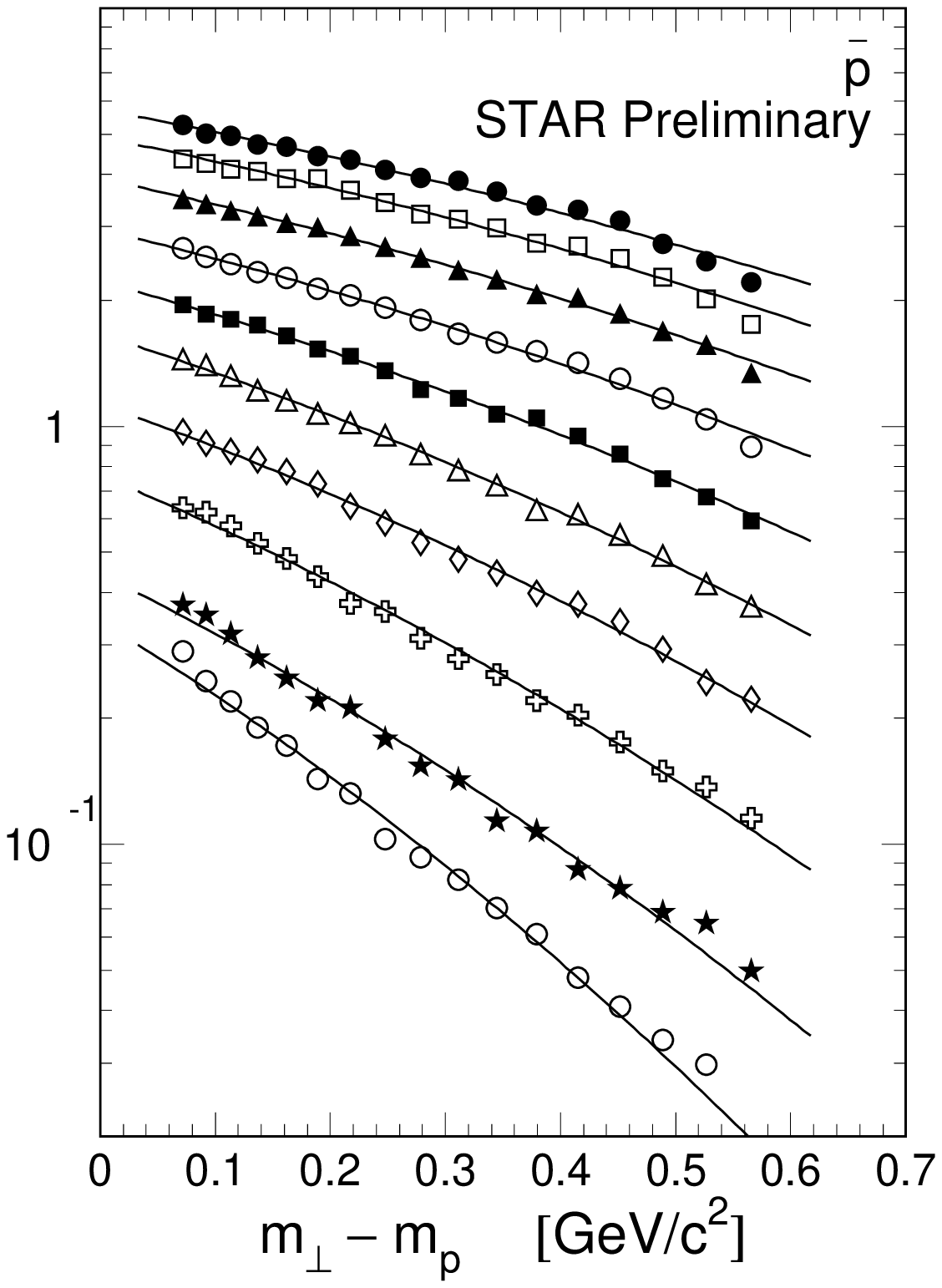}
\vspace {1cm}}
\caption{Preliminary mid-rapidity transverse mass spectra of 
$\pi^-$ (left), $K^-$ (middle), and $\overline{p}$ (right). 
The lowest spectrum is from 200 GeV p+p within $|{\rm y}|<0.25$ 
(scaled up by a factor of 5). 
The other spectra are from 200 GeV Au+Au within $|{\rm y}|<0.1$,
 in the order of decreasing centrality from top to bottom: 
5\% (most central), 5-10\%, 10-20\%, 20-30\%, 30-40\%, 40-50\%, 
50-60\%, 60-70\%, and 70-80\%. 
The pion spectra are corrected for weak decays and muon 
contaminations. 
The kaon and antiproton spectra are inclusive.}
\label{fqwang_fig1}
\end{figure}

Transverse momentum spectra of identified particles reflect 
the system properties in its final stage of thermal kinetic freeze-out, 
when all ellastic interactions between the constituents stop.
The essential parameters extracted from the data are the freeze-out
temperature and velocity of the radially expanding fireball.
Figure~\ref{fqwang_fig1} shows the mid-rapidity $\pi^-$, $K^-$, and 
$\overline{p}$ spectra for nine centrality bins in 200~GeV Au+Au 
collisions~\cite{Barannikova:2002qw}. 
Systematic errors on the spectra are estimated 
to be 10\%.
 The $\pi^+$ and $K^+$ spectra are similar to the $\pi^-$ and $K^-$ spectra, 
respectively. All the spectra are similar to the corresponding ones at 
130~GeV~\cite{Adler:2001aq,Adler:2002wn}. 
We fit the pion spectra with a Bose-Einstein function, 
the kaon spectra with a $m_\perp$ exponential function, and the 
antiproton spectra with a $p_\perp$ Gaussian function, 
to extract the mean transverse momenta, 
$\langle p_\perp \rangle$, and the $dN/d{\rm y}$ yields. 
The fit results are superimposed in Figure~\ref{fqwang_fig1}. 
The measured yields are about
 65\% of the extrapolated $dN/d{\rm y}$ for pions, 50-65\% for kaons, 
and 50-75\% for antiproton, respectively. 

\begin{figure}[htb]
\centerline{
\epsfxsize=0.45\textwidth\epsfbox[20 300 520 540]{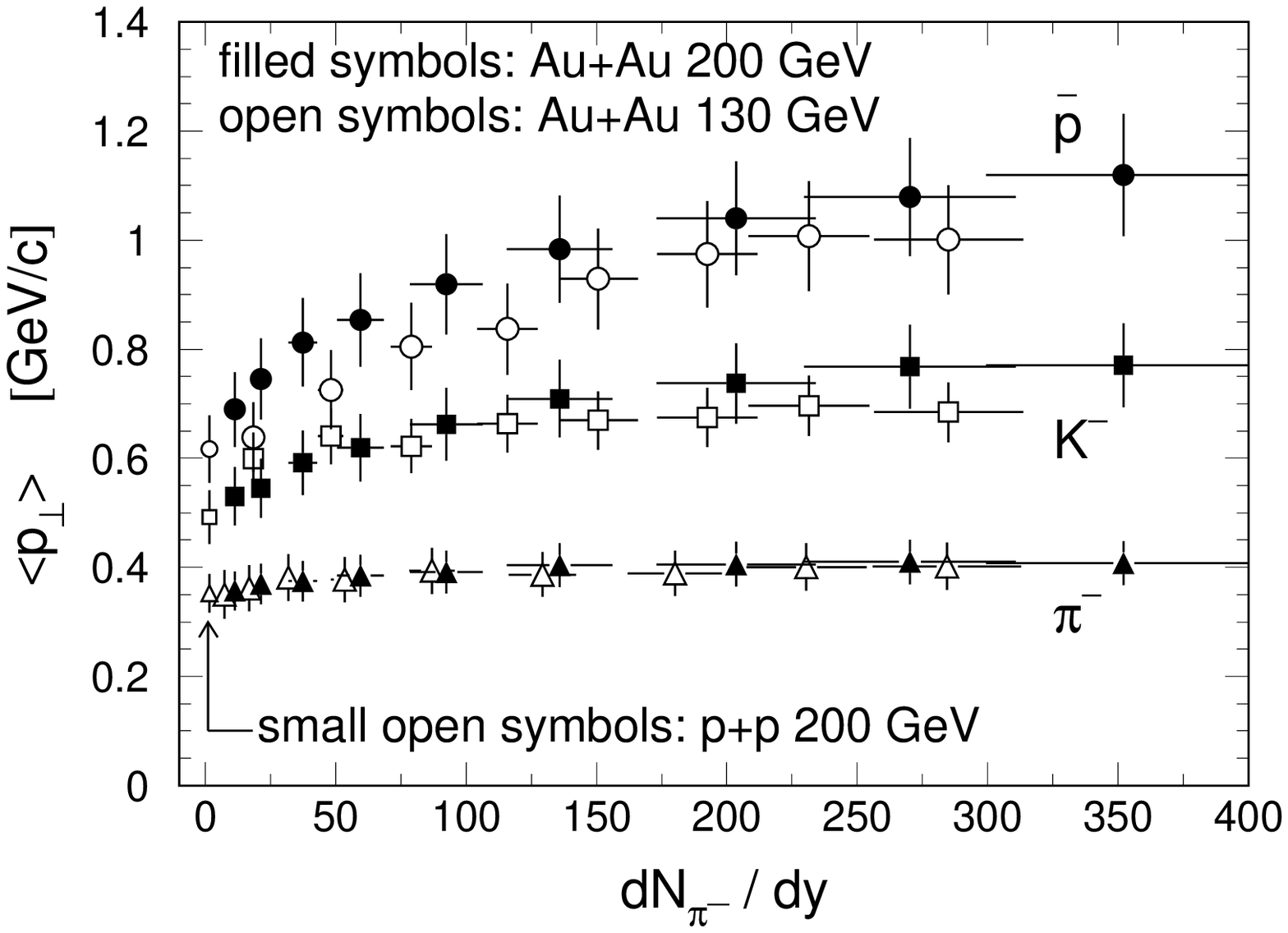}
\epsfxsize=0.45\textwidth\epsfbox[20 165 520 405]{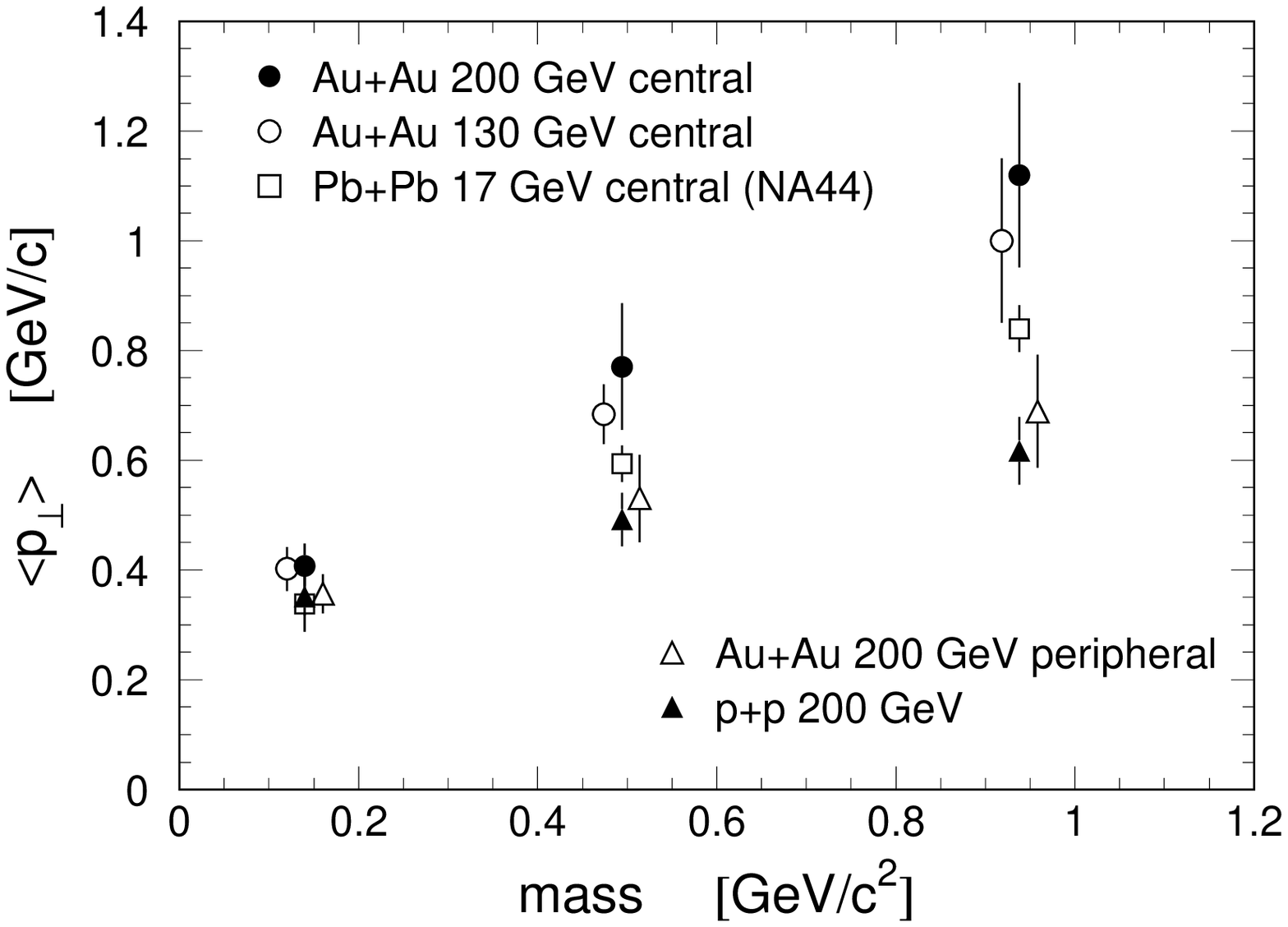}
\vspace {1cm}}
\caption{Left: Extracted $\langle p_\perp \rangle$ 
versus centrality for $\pi^-$ (triangles), $K^-$ (squares), and $\overline{p}$ 
(circles) from STAR. 
Right: Extracted $\langle p_\perp \rangle$ versus particle mass 
for five collision systems.}
\label{fqwang_fig2}
\end{figure}

Figure~\ref{fqwang_fig2} (left) shows the extracted 
$\langle p_\perp \rangle$ of $\pi^-$, $K^-$, and $\overline{p}$ 
as a function of the mid-rapidity $\pi^-$ multiplicity density, 
$dN_{\pi^-}/d{\rm y}$, used as a measure of the collision centrality. 
A systematic increase with centrality is observed in the kaon and 
antiproton $\langle p_\perp \rangle$, 
consistent with collective transverse radial flow being 
built up in non-peripheral collisions.

Figure~\ref{fqwang_fig2} (right) shows 
$\langle p_\perp \rangle$ versus particle mass for five selected systems. 
The p+p and most peripheral A+A results indicate no 
transverse radial flow; the increase in $\langle p_\perp \rangle$ 
with mass in these systems is due to a trivial mass effect. 
The central collision results deviate from p+p and are 
consistent with transverse radial flow, which appears to be stronger at RHIC
 than SPS. 

A blast wave fit to the central
$\pi^-$, $K^-$, and $\bar{p}$~spectra provides a transverse radial flow velocity
of $\langle\beta_{T}\rangle$$\simeq$0.55$c$ at 130 GeV, and $\sim$0.60$c$ at
200 GeV, with $c$ the speed of light. The fit kinetic
freezeout temperature parameters show at most a mild decrease
going from SPS ($T_{\rm fo}$$\simeq$110 MeV) to RHIC
($T_{\rm fo}$$\simeq$100 MeV) energies.

\section{Hard Physics}
High \pT\ hadrons are produced in the initial collisions of incoming
partons with large momentum transfer. Hard scattered partons fragment
into a high energy cluster (jet) of hadrons.
In elementary $e^+e^-$ and $pp$ collisions, jet cross sections and single 
particle spectra at high transverse momentum are well described over a 
broad range of collision energies by perturbative QCD (pQCD).
 Partons 
propagating through a dense system may interact with the surrounding medium
radiating soft gluons at a rate proportional to the energy density
of the medium. The measurements of radiative energy loss (jet quenching)
in dense matter
may provide a direct probe of the energy density\cite{Baier:2000mf}.

\subsection{High $p_T$: Suppression of Inclusive Spectra}

Hadrons from jet fragmentation may carry a large fraction
of jet momentum (leading hadrons). In the absence of nuclear medium
effects, the rate of hard processes should scale with the number of
binary nucleon-nucleon collisions. The yield of leading hadrons measured
in Au+Au collisions at \sqrtsNN=130 GeV 
has been shown to be significantly suppressed\cite{Adler:2002xw},
indicating substantial in-medium interactions. 
The high statistics 200 GeV data extended the measurements of hadron
spectra to $p_T=$12 GeV/c~\cite{Adams:2003kv}.

\begin{figure}
\centering
\mbox{
\subfigure{\includegraphics[height=0.43\textwidth]{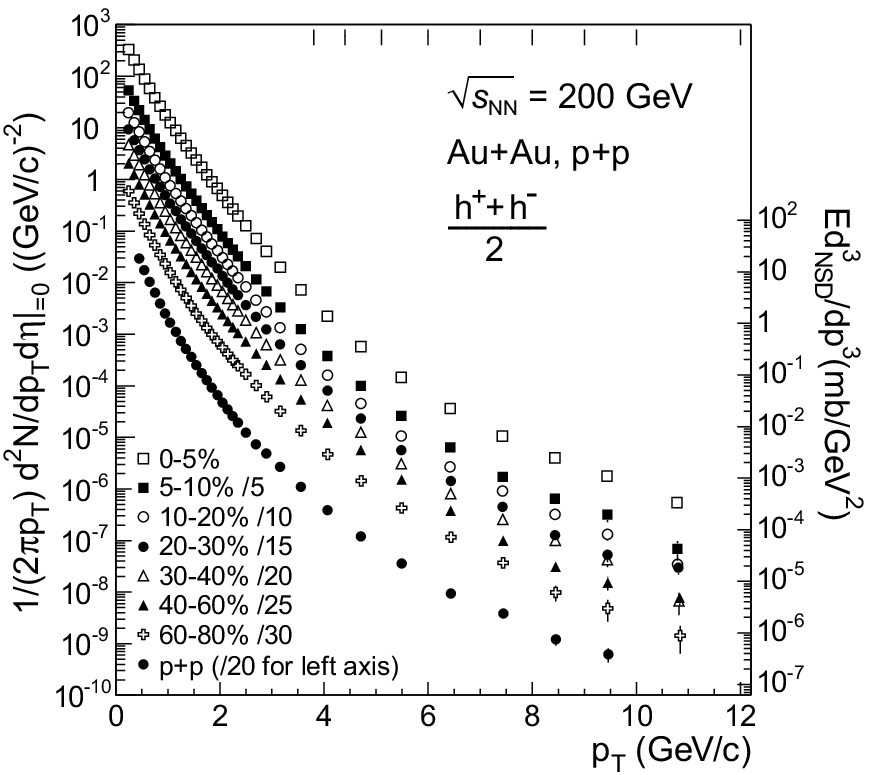}}
\subfigure{\includegraphics[height=0.43\textwidth]{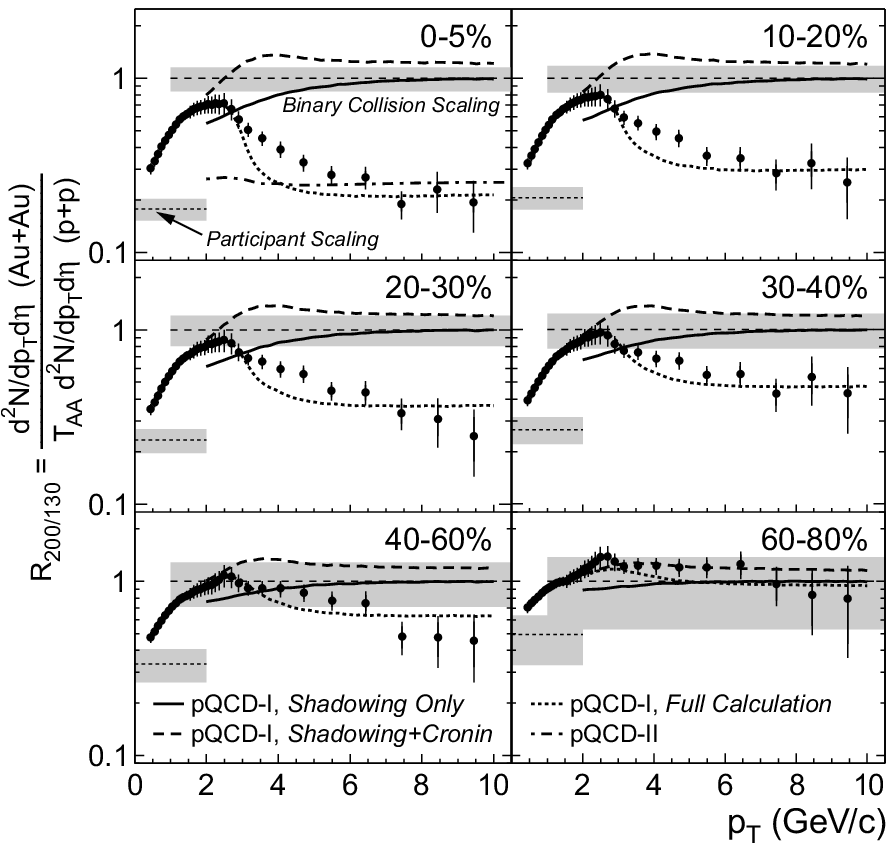}}
}
\caption{Left: Inclusive invariant \pT\ distributions of \hphm\ for centrality-selected Au+Au and p+p non-singly diffractive interactions. Right: \RAA\ (Eq.~\ref{RAA}) for \hphm\ in $|\eta|$\lt0.5, for
centrality-selected Au+Au spectra relative to the measured p+p spectrum. The horizontal dashed lines show Glauber model
expectations for scaling of the yield with mean number of binary collisions 
\NbinaryMean\ or mean number of participants
\NpartMean, with the grey bands showing their respective uncertainties summed in
with the p+p normalization uncertainty.
\label{spectra}}
\end{figure}

Figure~\ref{spectra} (left) shows inclusive invariant \pT\ distributions of
charged hadrons within $|\eta|$\lt0.5 for Au+Au and p+p collisions at
\sqrtsNN=200 GeV. The centrality-selected Au+Au spectra are shown for 
percentiles of \sigmaAAgeom, with 0-5\% indicating the most central collisions.
The measured hadron yields steeply decrease with increasing transverse momentum.
Modification of inclusive spectra by nuclear effects is measured by
comparison to a nucleon-nucleon (NN) reference via the
nuclear modification factor:
\begin{equation}
\label{RAA}
\RAA=\frac{d^2N^{AA}/d{\pT}d\eta}{\TAA{d}^2\sigma^{NN}/d{\pT}d{\eta}}\ ,
\end{equation}
\noindent
where \TAA=\NbinaryMean/\sigmaNNinel\ from a Glauber calculation
accounts for the nuclear collision
geometry. 
In the absence of nuclear effects, 
\RAA\ is expected to be unity.
Figure~\ref{spectra} (right) shows \RAA\ at \sqrtsNN=200 GeV for
centrality-selected Au+Au spectra relative to the measured p+p
spectrum.  For \pT\lt6 GeV/c, \RAA\ is similar to that observed at
\sqrtsNN=130 GeV~\cite{Adler:2002xw}, though in the present case
the NN reference and Au+Au spectra are measured at the same energy and
acceptance. Hadron production for 6\lt\pT\lt10 GeV/c is suppressed by
a factor of 4-5 in central Au+Au relative to p+p collisions.
The data are compared to two calculations based on hard parton scattering
evaluated via perturbative QCD (pQCD-I \cite{Wang:2003mm} and pQCD-II \cite{Vitev:2002pf}).
Both pQCD models for Au+Au collisions incorporate nuclear
shadowing of initial-state parton densities, the Cronin effect \cite{Antreasyan:cw},
and partonic energy loss, but with different formulations. Neither 
pQCD calculation includes non-perturbative effects that generate particle 
species-dependent differences for \pT$<5$ GeV/c~\cite{Vitev:2001zn}.
Figure~\ref{spectra} (right) also shows the full pQCD-I calculation and the
influence of each nuclear effect. The Cronin enhancement and shadowing
alone cannot account for the large suppression, which is reproduced
only if partonic energy loss in dense matter is included.

\subsection{High $p_T$: Elliptic Flow}
 The
fragmentation products of high energy partons that have propagated 
through the azimuthally 
asymmetric
system generated by non-central collisions may exhibit azimuthal
anisotropy due to energy loss and the azimuthal dependence of the path
length \cite{Wang:2000fq,Gyulassy:2000gk}.
The azimuthal anisotropy of final
state hadrons in non-central collisions is quantified by the
coefficients of the Fourier decomposition of the azimuthal particle
distributions, with the second harmonic coefficient $v_2$ referred to
as elliptic flow \cite{Voloshin:1994mz}. The elliptic flow measurements for $p_T<2$ GeV/c 
agree in detail with hydrodynamic calculations\cite{Ackermann:2000tr}.
\begin{figure}[t]       
\centering      
\mbox{
\subfigure{\includegraphics[width=0.46\textwidth]{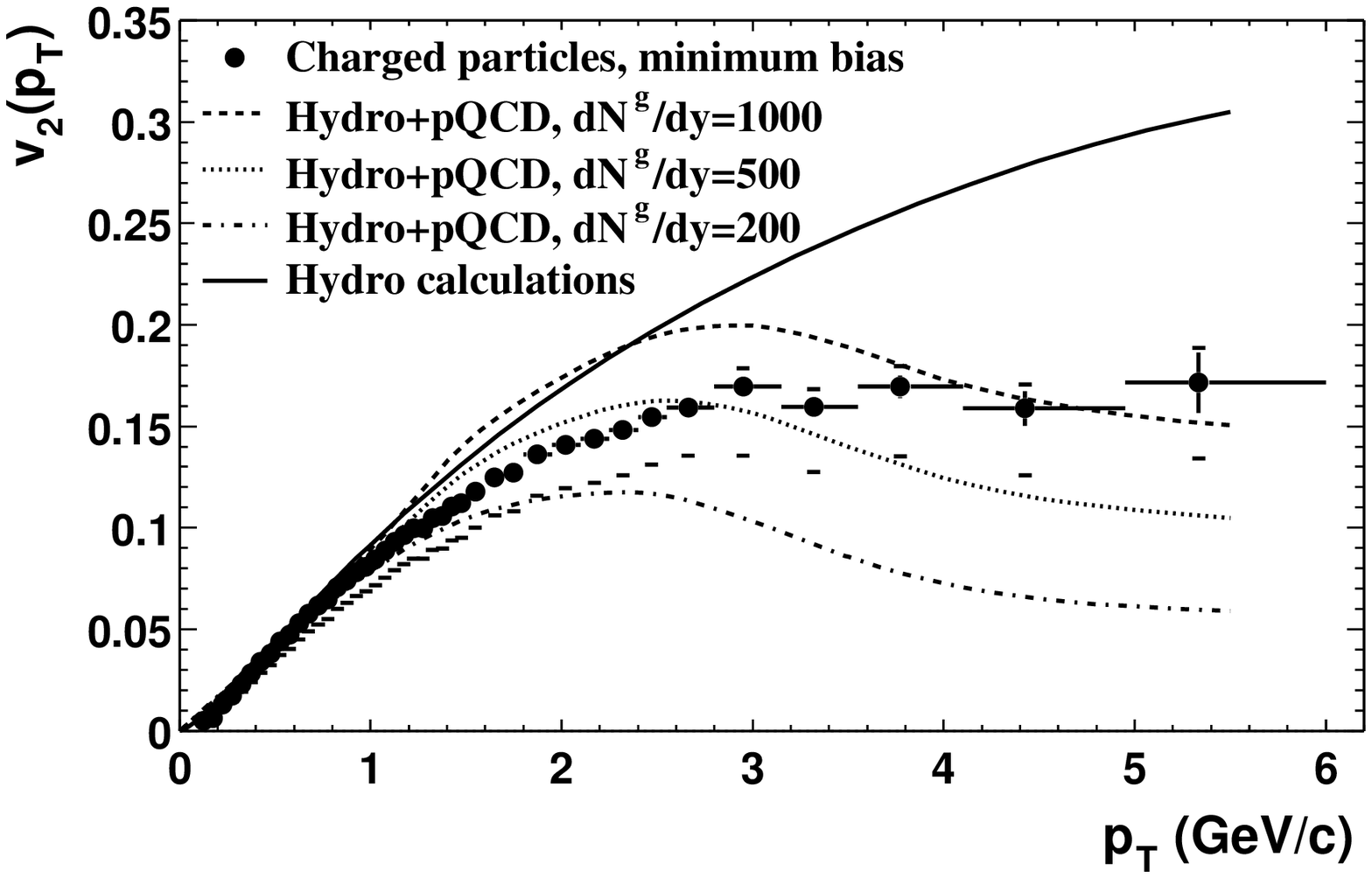}}
\subfigure{\includegraphics[width=0.44\textwidth]{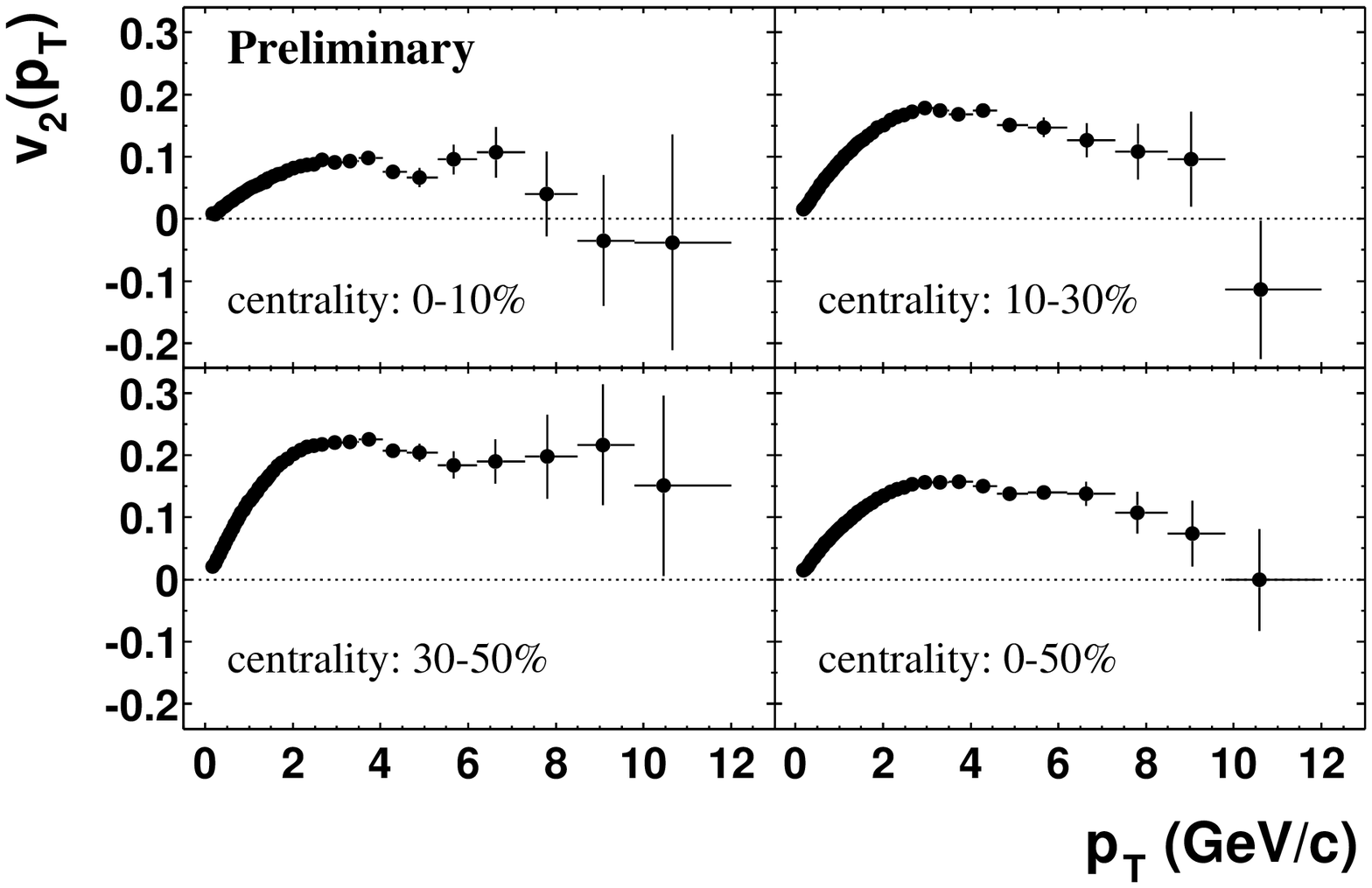}}
}
\caption{Elliptic flow for charged hadrons. Left panel: $v_2(p_T)$ for 
minimum bias
 at \sqrtsNN=130 GeV compared to hydro and pQCD calculation.
Right panel: $v_2(p_T)$ for different centralities at \sqrtsNN=200 GeV.}
\label{v2}
\end{figure}
Figure~\ref{v2} (left panel) shows the elliptic flow $v_2$ as a function
of $p_T$ for Au+Au collisions at 130 GeV \cite{Adler:2002ct}. Elliptic
flow rises almost linearly with transverse momentum up to 2 GeV/c,  
behavior that is 
well described by hydrodynamic calculation. Above $p_T\sim 2$ GeV/c,
$v_2(p_T)$ deviates from a linear rise and saturates for $p_T>3$ GeV/c.
The azimuthal anisotropies measured at $p_T=4-6$ GeV/c are in
qualitative agreement with the pQCD calculations including 
energy loss\cite{Gyulassy:2000gk}. Figure~\ref{v2} (right panel) 
shows the centrality
dependence of $v_2(p_T)$ measured at \sqrtsNN=200 GeV~\cite{Filimonov:2002xk}.
$v_2$ remains finite for non-central collisions, exhibiting a decrease from 
the saturation level
at the highest measured $p_T$ for the more central events.
It is expected
that the azimuthal anisotropy will vanish in the limit of very high $p_T$.
However, at present
the measured values of $v_2$ contain 
a non-flow component which at high \pT\ comes
from intra-jet correlations.
A quantitative understanding of this effect at the highest \pT\,
especially for the most peripheral and central collisions
is still needed.


\subsection{High $p_T$: Near-angle and Back-to-Back Hadron Correlations}

The large multiplicities in nuclear collisions make full jet reconstruction
impractical. Correlations of high \pT\ hadrons in pseudorapidity and azimuth
allow the identification of jets on a statistical basis. 
First hints of jets at RHIC came from the two-particle azimuthal distributions
at \sqrtsNN=130 GeV \cite{Adler:2002ct}.
Similar analysis performed for \sqrtsNN=200 GeV \cite{Adler:2002tq}
directly shows that hadrons at $p_T>3-4$ GeV/c result from the fragmentation
of jets.

\begin{figure}[ht]
\centering
\mbox{
\subfigure{\includegraphics[width=0.49\textwidth]{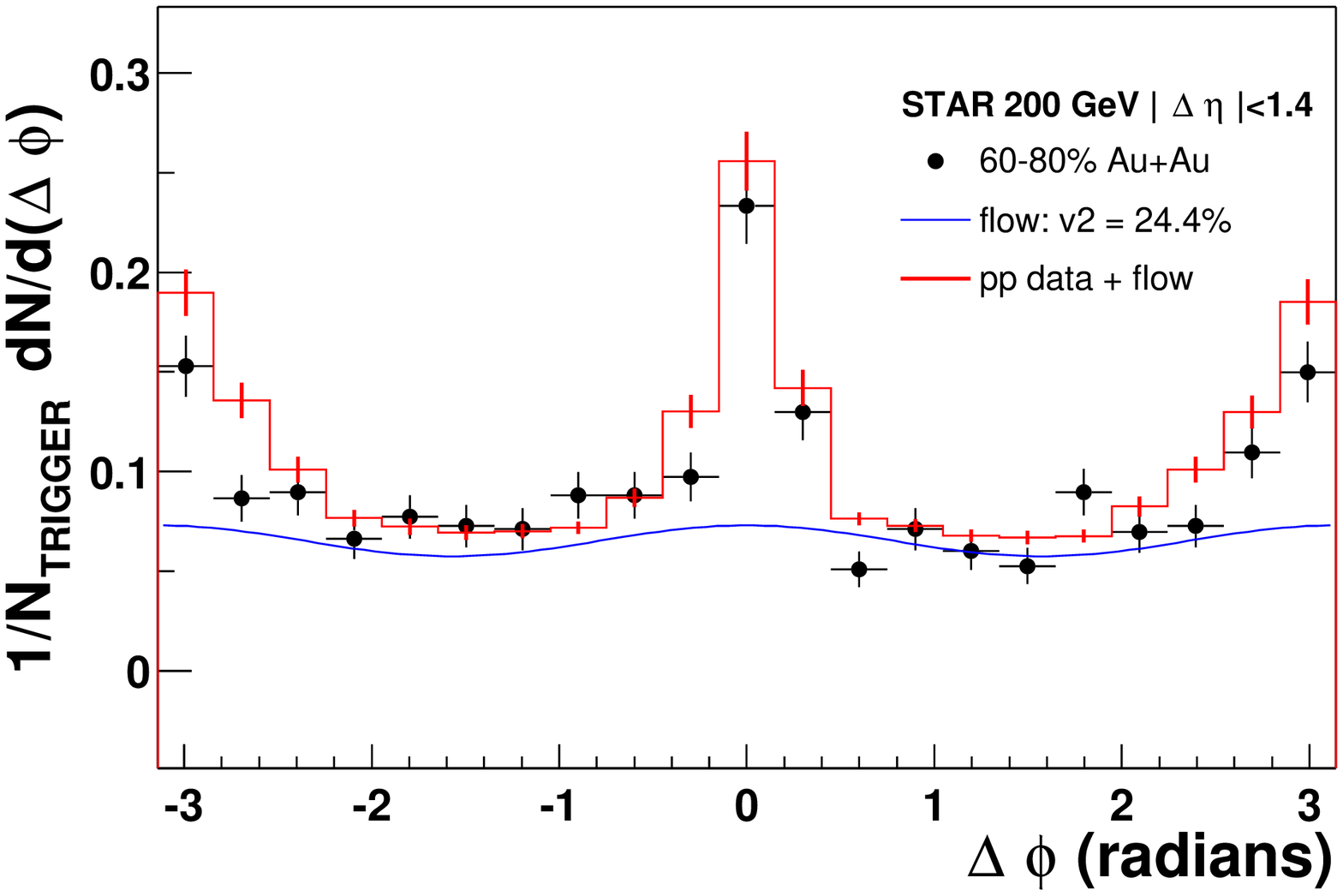}}
\subfigure{\includegraphics[width=0.49\textwidth]{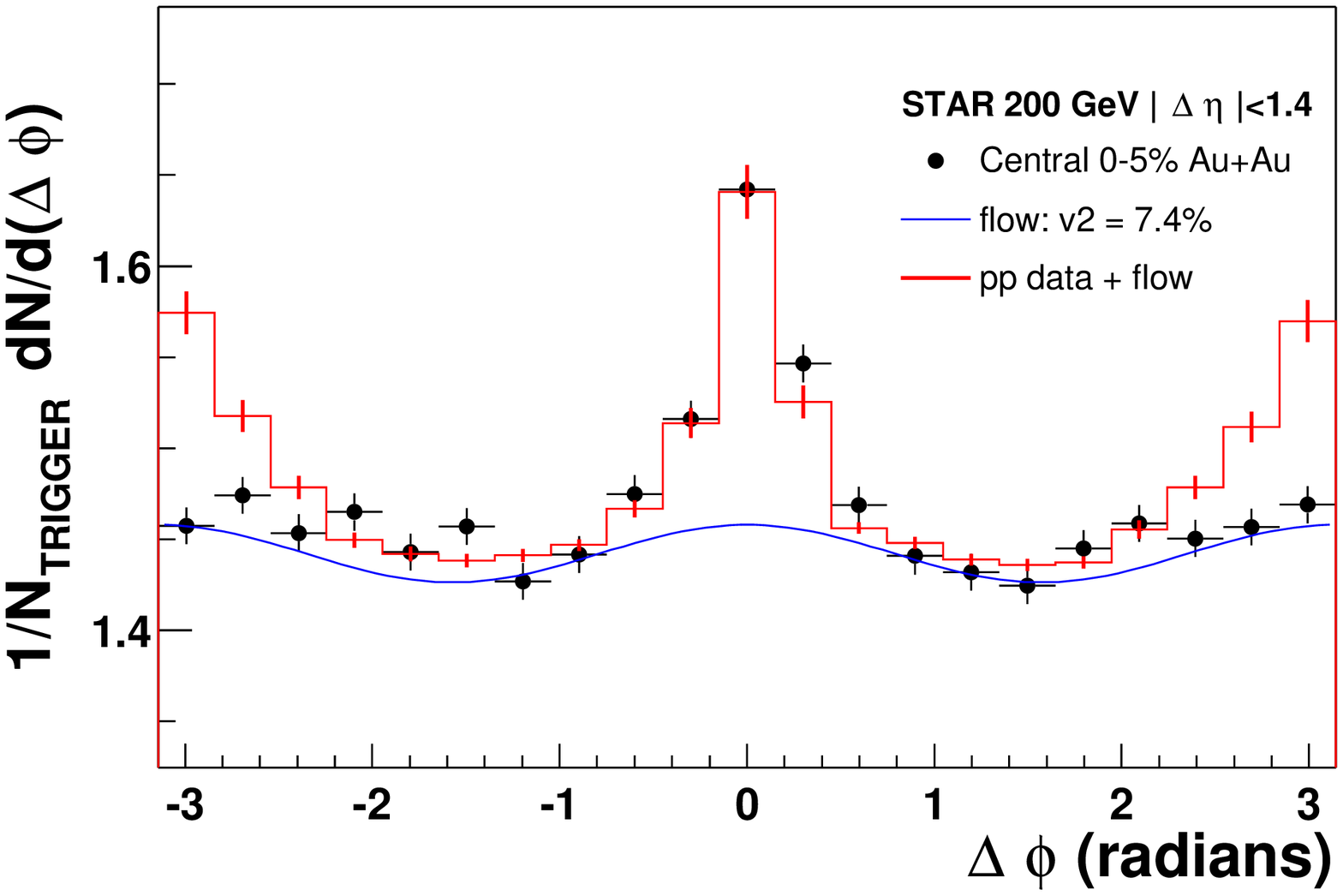}}
}
\caption{Azimuthal distributions of high \pT\ hadron pairs in Au+Au compared to p+p plus elliptic flow (STAR). Left panel: peripheral collisions. Right panel: central collisions.\label{jets}}
\end{figure}

\begin{figure}[tb]
\begin{center}
\includegraphics[height=5cm]{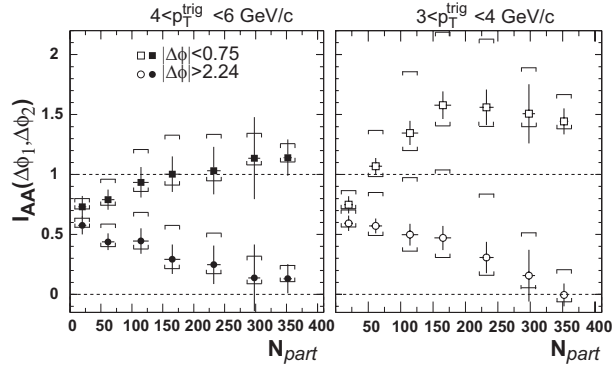}
\end{center}
\caption{Ratio of Au+Au and p+p (Equation \protect{\ref{ratioeqn}}) 
for small-angle (squares, $|\Delta \phi|<0.75$ radians) and 
back-to-back (circles, $|\Delta \phi|>2.24$ radians) azimuthal regions 
versus number of participating nucleons 
for trigger particle intervals $4<p_T^{trig}<6$ GeV/c (solid) and
$3<p_T^{trig}<4$ GeV/c (hollow). The horizontal bars indicate the 
dominant systematic error (highly correlated among points) due
to the uncertainty in $v_2$.}
\label{suppress}
\end{figure}

Events with at least one large transverse momentum hadron
($4<p_T^{trig}<6$ or $3<p_T^{trig}<4$ GeV/c), defined to be a {\it trigger}
particle, are used in this analysis. The trigger particles are paired
with {\it associated} particles with 2~GeV/c~$<p_T<p_T^{trig}$.
An overall azimuthal pair distribution per trigger particle is then
constructed:	
\begin{eqnarray}
D(\Delta \phi) \equiv \frac{1}{N_{trigger}}\frac{1}{\epsilon} \int d\Delta
\eta N(\Delta \phi, \Delta \eta),
\end{eqnarray}
where $N_{trigger}$ is the
observed number of tracks satisfying the trigger requirement,
and $\Delta \phi$, $\Delta \eta$ are the azimuthal and pseudo-rapidity 
separations between the trigger and associated particles. 
The efficiency $\epsilon$ 
for finding the associated particle is evaluated by embedding simulated
tracks in real data. 
Figure~\ref{jets} shows the azimuthal angular distribution between pairs
of high \pT\ hadrons for the peripheral and most central Au+Au collisions
at \sqrtsNN=200 GeV. 
The strength of near-side correlations (near $\Delta \phi$ = 0) for both 
centralities is consistent with that measured in p+p collisions.
This indicates that
the same mechanism (hard parton scattering and fragmentation)
is responsible for high transverse momentum particle production in p+p
and Au+Au collisions. 
The away-side (back-to-back) correlations in peripheral Au+Au
collisions may be described by an incoherent superposition of jet-like
correlations measured in p+p and elliptic flow. However, back-to-back
jet production is strongly suppressed in the most central Au+Au collisions.
This indicates a substantial interaction as the hard-scattered partons
or their fragmentation products traverse the medium.
The ratio of the measured Au+Au correlation excess 
relative to the p+p correlation is:
\begin{eqnarray}
\lefteqn{I_{AA}(\Delta \phi_1,\Delta \phi_2) = } \nonumber\\
& \frac{\int_{\Delta \phi_1}^{\Delta \phi_2} d(\Delta \phi) [D^{\mathrm{AuAu}}- B(1+2v_2^2 \cos(2 \Delta \phi))]}{\int_{\Delta \phi_1}^{\Delta \phi_2} d(\Delta \phi) D^{\mathrm{pp}}}.
\label{ratioeqn}
\end{eqnarray}
The ratio can be plotted as a function of
the number of participating nucleons ($N_{part}$). 
 $I_{AA}$ is measured for both the small-angle 
 ($|\Delta \phi|<0.75$ radians) and back-to-back ($|\Delta \phi|>2.24$ radians) regions. The ratio should be unity
if the hard-scattering component of
Au+Au collisions is simply a superposition of p+p collisions unaffected
by the nuclear medium.  
These ratios are given in Figure~\ref{suppress} for the trigger 
particle momentum ranges indicated.
For the most peripheral bin (smallest $N_{part}$), both the 
small-angle and back-to-back
correlation strengths are suppressed, which may be an indication
of initial state nuclear effects such as shadowing of parton distributions
or scattering by multiple nucleons, or
may be indicative of energy loss in a dilute medium.
As $N_{part}$ increases, 
the small-angle correlation strength increases, with 
a more pronounced increase 
for the trigger particles with lower $p_T$ threshold.  
The back-to-back correlation strength, above background from elliptic flow,
decreases with increasing $N_{part}$ and 
is consistent with zero for the most central collisions.

\section{Summary}
At RHIC, a system with low net-baryon density at mid-rapidity is produced.  
2/3 of the baryons come from baryon-antibaryon pair production, while 1/3
of mid-rapidity net baryons come from the initial nuclei.  
The system
undergoes chemical freeze-out at $T_{ch}\approx 175$ MeV and $\mu_{b}\approx 25-50$ MeV.  
The single particle $p_T$ spectra
suggest that the system undergoes further
elastic re-scattering until final 
freeze-out at $\langle\beta_T\rangle\approx 0.50-0.6c$ and a 
kinetic freeze-out temperature $T_{fo}\approx 100$ MeV.

The striking phenomena that have been
observed at high $p_T$ in nuclear collisions at RHIC: 
strong suppression of the
inclusive hadron yield in central 
collisions, large elliptic
flow which saturates at $p_T>3$ GeV/c, and disappearance of back-to-back jets,
 are all consistent
with a picture in which observed hadrons at $p_T>3-4$ GeV/c are fragments of
hard scattered partons, and partons or their fragments are strongly
scattered or absorbed in the nuclear medium. The observed hadrons therefore
result preferentially from hard-scattered partons generated on the periphery of the
reaction zone and heading outwards.
 These observations appear consistent
with large energy loss in a system that is opaque to the propagation
of high-momentum partons or their fragmentation products. 
The upcoming results from d-Au collisions at \sqrtsNN=200 GeV
will help in determining whether the observed effects are due to 
the final state interactions.

\end{document}